\documentclass{lmcs} 
\usepackage{graphicx}

\usepackage{times}
\usepackage{comment}
\usepackage{pstricks}
\usepackage[latin9]{inputenc}
\usepackage{amstext}
\usepackage{tikz}
\usepackage{graphicx}
\usepackage{amssymb}
\usepackage{multirow}
\usepackage{caption}
\usepackage{enumitem}

\usepackage{algorithm}
\usepackage{float}
\usepackage{algorithmic}
\usepackage{tabularx}

\usepackage{multirow}
\usepackage{verbatim}
\usepackage{rotating}
\usepackage{lastpage}
\usepackage{lmodern}
\usepackage{microtype}
\usepackage{latexsym}
\usepackage{lscape}
\usepackage{array}
\usepackage{xcolor}
\usepackage{amsmath}
\usepackage{tikz}
\usetikzlibrary{positioning,shapes,arrows,backgrounds}
\tikzset{
	events/.style={ellipse, draw, align=center},
}
\newcommand{\K}{{\mathcal{K}}}
\newcommand{\T}{{\mathcal{T}}}
\newcommand{\A}{{\mathcal{A}}}
\newcommand{\R}{{\mathcal{R}}}
\newcommand{\C}{{\mathcal{C}}}

\newcommand{\I}{{\mathcal{I}}}

\newcommand{\KB}[2]{\left<#1,#2\right>}
\newcommand{\dllite}{{{DL-Lite}}}
\newcommand{\ie}{\textit{i.e.,}}
\newcommand{\eg}{\textit{e.g.}}
\newcommand{\resp}{\textit{resp.}}
\newcommand{\wrt}{w.r.t.}
\newcommand{\kb}{KB}
\newcommand{\finexample}{\hfill $\blacksquare$}
\floatname{algorithm}{Algorithm}
\newcommand{\isa}{\sqsubseteq}


\usepackage{hyperref}
\theoremstyle{plain} 

\def\eg{{\em e.g.}}

\begin{document}
	
	\title{Recursive Algorithms To  Repair prioritized and inconsistent \textit{DL-Lite} knowledge Base}
	
	\author{Ghassen Hamdi}	
	\address{MARS Research Laboratory, University of Sousse, Tunisia}	
	\email{hamdighassan@gmail.com}  
	
	\author{TELLI Abdelmoutia}	
	\address{ Biskra University, Algeria}	
	\email{a.telli@univ-biskra.dz}  
	
	\author{Mohamed Nazih Omri}	
	\address{MARS Research Laboratory, University of Sousse, Tunisia}	
	\email{mohamednazih.omri@fsm.rnu.tn}  
	
	
	
	
	
	\begin{abstract}  
		\noindent The inconsistency in prioritized knowledge base is because the assertions (ABoxes) come from several  sources with different levels of reliability. We  introduce the handling of this  inconsistency problem  to query inconsistent \textit{DL-Lite} knowledge bases.  In the literature, firstly, repair all the inconsistent assertions of the \textit{DL-Lite}'s inconsistent knowledge base. Then, interrogate it. However, our algorithm, on proceeds directly with an interrogation of the knowledge base in order to recover an exhaustive list of answers to a given query. In a second time, to repair the answers of this list. The novelty of our article is the proposition of a recurring function that calculates the rank of coherence in order to manage the inconsistencies in the set of responses. This strategy allowed us to reduce execution time compared to existing algorithms. The experimental study as well as the analysis of the results, which we carried out, showed that our algorithm is much more productive than the other algorithms since it gives the greatest number of answers while remaining the best from the point of view of the execution time. Finally, as shown in our experimental studies, they allow an efficient handling of inconsistency. Such facts make all the repairs suitable for \textit{DL-Lite}.
	\end{abstract}
	
	\maketitle
	
	\section{Introduction}\label{S:one}
	
	Description Logics (DLs) are formal frameworks for representing and reasoning with ontologies.  The DL knowledge base consists of: TBox as a terminological base represents the conceptual knowledge of a particular domain and ABox as an assertional base contains facts or assertions concerning particular individuals \cite{ref_DLHB2010}.
	
 Recently, Ontology Based Data Access (OBDA) \cite{ref_Lenzerini82,ref_DBLP08,ref_DBLP13} is collecting great attention as a new paradigm in which structured knowledge or the ontological view (\ie \; stored in a TBox) is used to provide better exploitation of assertions (\ie \; stored as an ABox) when querying them. A crucially important service provided by an OBDA system is query answering, which aims to calculate the answers of a query posed in terms of ontologies.


In many applications, assertions are provided by several potentially conflicting sources with different levels of reliability. In addition, a given source may have different sets of uncertain assertions that together form a prioritized or a stratified assertional base (i.e ABox). In the case, where the data is provided from multiple and unreliable sources that may be inconsistent. A major problem arises in the context of \textit{DL-Lite} query answering is how to deal with the case of inconsistency between ABox and TBox. Indeed, the TBox is generally verified and validated as long as the ABox will typically be larger and may have several modifications and thus may be in contradiction with the TBox.

Many works in the context of OBDA, inspired by database approaches (\eg\; \cite{ref_Arenas99,ref_Bertossi2011,ref_Chomicki2007}) or propositional logic approaches (\eg\; \cite{ref_BenferhatCDLP93,ref_BenferhatDPW97,ref_Nebel94}), deal with the problem of querying inconsistent DL \kb s by proposing several inconsistency-tolerant inferences, called semantics,  and were introduced for the lightweight description logic \textit{DL-Lite} \eg\; \cite{ref_lemboLRRS15}. Among these semantics, one can quote the AR and IAR semantics \cite{ref_LemboLRRS10} which are the most known and the most studied. These two semantics are based on the notion of maximal assertional repair, which is based on the notion of repair in the database domain or the maximal consistent subsets in the propositional logic setting \cite{ref_Nicholas1970,ref_Brewka89}.

The main contribution of this article is to develop a recursive function in order  to deal with the inconsistency of the answers to a query with respect to the TBox. Our experiments focus on the running time, the precision calculation, the recall, the F-measure properties and the productivity of these inferences strategies after running a query.

\indent The rest of this article is organized as follows: next Section \ref{rw} presents the related works. Section \ref{syn} provides a brief refresher on \textit{DL-Lite} and querying multiple prioritized sources. Section \ref{inc} explains the notion of inconsistency tolerant reasoning for assertions associated with the answers profile (Conflicts set repair answers, Free set answers, reparis answers, and consistency rank). Section \ref{poslin} introduces our proposed  strategies of answers profile repairs. Section \ref{exp} presents the experimental studies and Section \ref{conclusion} concludes the paper.


	\section{Syntax and Semantics of Prioritized \textit{DL-Lite} Knowledg Base}
	\label{syn}
\noindent Let $N_C$, $N_R$ and $N_I$, three pairwise disjoint sets of atomic concepts, atomic roles and individuals respectively. Let $A \in N_C$, $P \in N_R$. Let also  '$\neg$', '$\exists$' and '$^{-}$' three connectors  are used to define complex concepts and roles. \textit{DL-Lite} concepts are defined as follows:
\begin{eqnarray*}
	\label{eq:dll}
	\quad\quad R\longrightarrow\quad P\quad|\quad P^{-} & \quad  & E\longrightarrow\quad R\quad|\quad\neg R\\
	B\longrightarrow\quad A\quad|\quad K & \quad &  C\longrightarrow\quad B\quad|\quad\neg B
\end{eqnarray*}

A \textit{{\dllite}} \kb $\;$ is a pair $\K$=$\KB{\T}{\A}$ where $\T$ is called the TBox and $\A$ is called the ABox. A TBox includes a finite set of inclusion axioms on concepts and on roles respectively of the form inclusion assertions:  $B\sqsubseteq C$ (\resp \;  negative inclusion assertions $B\sqsubseteq  \neg C$) means that concept  $B$ is included in concept $C$ (\resp \;  concept  $B$ is not included in the concept $C$) and $R\sqsubseteq E$  (\resp \; $R \sqsubseteq  \neg E$) means that  role  $R$ is included in the role $E$ (\resp \;  role $R$ is not included in role $E$).

 The ABox contains a finite set of assertions (facts) of the form $A(a)$ and  $P(a,b)$ where  $A \in N_C$, $P \in N_R$  and  $a,b \in N_I$. The semantics is given in terms of interpretations $\I$=$(\Delta^{\I},.^{\I})$ which  consist of an non-empty domain $\Delta^{\I}$ and an interpretation function $.^{\I}$ that  assigns to each $a \in N_I$ an element $a^{\I} \in \Delta^{\I}$, to each $A \in N_C$ a subset $A^{\I} \subseteq \Delta^{\I}$ and to each $P \in N_R$ an  $P^{\I} \subseteq \Delta^{\I} \times \Delta^{\I}$. A TBox $\T$ is said to be incoherent if there exists a concept $C$ s.t $\forall \I$:$\I\models \T$, we have $C^{\I}$=$\emptyset$. A \textit{\dllite{}} \kb{} $\K$ is said to be inconsistent if it does not admit any model.

The prioritized profile is  a multiset of prioritized or stratified ABoxes denoted by $\K_P = \KB{\T}{P_s}$  where $\T$ is a standard \textit{DL-Lite} TBox and $P_s=\{L_1,. . .,L_m\}$ is a prioritized ABox profile. It is assumed that each ABox $L_i\in P_s$ is consistent with the ontology (TBox). In this case, each sets $L_i$ are called layers or strata, which each layer $L_i$ contains a set of assertions with the same level of priority $i$ and they are considered more reliable than those present in a layer $L_j$ when $j> i$. Accordingly, $L_1$ contains the most important assertions as long as $L_m$ contains the least important ones. 

A query is a first-order logic formula, denoted $q$=$\{(x) \:| \phi(x)\}$, where $(x)$=($x_1$,...,$x_n$) are free variables, $n$ is the arity of $q$ and atoms of $\phi(x)$ are of the form $A(t_i)$ or $P(t_i,t_j)$ with $A \in N_C$ and $P \in N_R$ and $t_i$, $t_j$ are terms, \ie{} constants of $N_I$ or variables.  When  $\phi(x)$ is of the form $\exists (y).conj(x,y)$ where $y$ are bound variables called existentially quantified variables, and $conj(x,y)$ is a conjunction of atoms of the form $A(t_i)$ or $P(t_i,t_j)$ with $A \in N_C$ and $P \in N_R$ and $t_i$, $t_j$ are terms, then $q$ is said to be a conjunctive query (CQ). An answer to a CQ $ q(x) \leftarrow conj(x,y)$ over a KB $\K=\KB{\T}{\A}$ is a non empty set of tuples $s = (s_1, \cdots , s_k) \in N^{\I} \times \cdots \times N^{\I}$ such that $\KB{\T}{\A} \models  q(s)$  $\KB{\T}{\A} \models  q(s)$.

Let $q(x)$ be a query, we consider $S_{P_s}=\{S_1,..., S_m\}$ a various sets of answers to a query $q(x)$ regarding the prioritized profile $P_s$ called the profile of sets of answers where each $S_i$ is the set of answers to $ q(x)$  \wrt{} $L_i$ for $1 \leq i \leq m $, defined as follows : $S_i = \{s \in N^\I \times \cdots \times N^\I :   \KB{\T}{L_i} \models  q(s) \}$. Certainty, when there is no answer to the query $q(x)$ with respect to $L_i$, $S_i$ = $\emptyset$.

\section{Inconsistency Tolerant Reasoning for Assertions
	Associated with the Answers Profile}
\label{inc}

In this section, we have assumed that each ABox is consistent with the TBox. Coping with inconsistency can be done by first computing the set of consistent subsets of assertions  associated to a set of answers to a given query, called repairs answers.

\subsection{Conflicts set Answers}
The conflicts sets answers represent a minimal inconsistent subset $\C$ of the assertions associated to $S_{P_s}$ such that $\langle\T,\C\rangle$ is inconsistent.
	\label{def3}
	
	Let $\K_P{=}\KB{\T}{P_s}$ be a prioritized \textit{DL-Lite} \kb $\;$ with: $P_s=\{L_1,\ldots,L_m\}$, $q(x)$ be a  query, $S_{P_s}=\{S_1,..., S_m\}$ a set of answers to a query $q(x)$ with respect to $P_s$, $q_{P_s} = (q_{L_1},\ldots,q_{L_m})$ is the set of assertions associated to $S_{P_s}$ such that $q_{L_i} = \{q(s) \, : \,  s \in S_{i}\}$. A subset $\C \subseteq q_{P_s}$ is said to be a conflicts sets answers iff $\langle\T,\C\rangle$ is inconsistent and $\forall f \in \C$, $\langle\T,\C \setminus \{f\}\rangle$ is consistent.

\subsection{Free Set Answers}
We  denote by $\textit{free}(q_{P_s})$ the set of assertions belong to $q_{P_s}$ that are not responsible for conflicts in $\langle\T,q_{P_s}\rangle$.
	\label{def4}

	Let $\K_P{=}\KB{\T}{P_s}$ be a prioritized \textit{DL-Lite} \kb $\;$ with: $P_s=\{L_1,\ldots,L_m\}$, $q(x)$ be a  query, $S_{P_s}=\{S_1,..., S_m\}$ a set of answers to a query $q(x)$ with respect to $P_s$, $q_{P_s} = (q_{L_1},\ldots,q_{L_m})$ is the set of assertions associated to $S_{P_s}$ such that $q_{L_i} = \{q(s) \, : \,  s \in S_{i}\}$. A $\textit{free}$ assertion $f \in q_{P_s}$ is said to be $\textit{free}$ if and only if $\forall c \in \C(q_{P_s}):f \notin c$. This notion of $\textit{free}$ elements is formerly proposed by \cite{ref_BenferhatDP92} in a propositional logic setting.


\subsection{Repairs  Answers}

A subset $\R_A \subseteq (q_{L_1} \cup \ldots \cup q_{L_m} )$ is said to be a repair answers if $\langle\T,\R_A \rangle$ is consistent and $\R_{A}$ is said to be a maximally inclusion-based repair answers of $q_{P_s}$, denoted by $MAR_A$, if $\langle\T,\R_A \rangle$ is consistent and $\forall \R^{'}_A \subseteq (q_{L_1} \cup \ldots \cup q_{L_m}) : \R_A \subsetneq \R^{'}_A, \R^{'}_A$ is inconsistent. According to this definition of $MAR_A$, adding any assertion $f$ from $(q_{L_1} \cup \ldots \cup q_{L_m}) \setminus \R_A$ to $\R_A$ implies the inconsistency of $\langle\T,\R_A \cup \{f\}\rangle$. Furthermore, the maximality in $MAR_A$ is used in the sense of set inclusion. We denote by $MAR_A (q_{P_s})$ the set of $MAR_A$ of $q_{P_s}$ with respect to $\T$. The definition of $MAR_A$ is similar to that defined in \cite{ref_LemboLRRS10}. Using the concept of repair answers, the treatment of inconsistency in flat \textit{DL-Lite} \kb s can be done by applying standard query answering either using the whole set of repairs answers (universal entailment or AR-entailment \cite{ref_LemboLRRS10}) or only using one repair answers.
A repair answers is defined as follows:
	\label{def2}

	Let $\K_P{=}\KB{\T}{P_s}$ be a prioritized \textit{DL-Lite} \kb $\;$ with: $P_s=\{L_1,\ldots,L_m\}$, $q(x)$ be a query, $S_{P_s}=\{S_1,..., S_m\}$ a set of answers to a query $q(x)$ with respect to $P_s$, $q_{P_s} = (q_{L_1},\ldots,q_{L_m})$ is the set of assertions associated to $S_{P_s}$ such that $q_{L_i} = \{q(s) \, : \,  s \in S_{i}\}$.


\subsection{Consistency Rank}

Generally, the checking of consistency degree and several inference services can be done with standard DLs reasoning services through consistent subsets of DL knowledge base has been explained in \cite{qi2007a} and \cite{qi2007b}. Clearly, the computing of inconsistency degree comes down to perform a dichotomie search in standard DL, and it is closely related to the method proposed in \cite{ref_Didier94} for computing inconsistency degrees of a possibilistic propositional knowledge base.

This notion of consistency rank defined is inspired by the degree of inconsistency used in the possibilistic logic where the degrees are encoded using values in the unit interval $[0, 1]$. It is counterpart of the algorithm proposed in \cite{ref_Didier94} (\resp \; \cite{qi2007a}) in the propositional logic (\resp \; description 	logic) setting.

	\label{def5}
	Let $\K_P{=}\KB{\T}{P_s}$ be a prioritized \textit{DL-Lite} \kb $\;$ with: $P_s=\{L_1,\ldots,L_m\}$, $q(x)$ be a query, $S_{P_s}=\{S_1,..., S_m\}$ a set of answers to a query $q(x)$ with respect to $P_s$, $q_{P_s} = (q_{L_1},\ldots,q_{L_m})$ is the set of assertions associated to $S_{P_s}$ such that $q_{L_i} = \{q(x) \, : \,  \vec{s} \in S_{i}\}$. The consistency rank of $q_{P_s}$, denoted by : $CnsRank(q_{P_s})$ is defined as follows:
	$$CnsRank(q_{P_s}) = max\{i \; where \; \langle\T,(q_{L_1},\ldots,q_{L_i})\rangle \; is \; consistent \} $$  

We propose the recursive Function \ref{fc} which calculate the consistency rank of $q_{P_s}$. Then, we will use it in all our proposed strategies of computing a consistent assertions associated to the answers. Formally, this recursive function is faster than that sequential one proposed in \cite{ref_Didier94}.
\begin{algorithm}[H]
	\caption{$\textit{CnsRank}$ $ (\T,(q_{L_i},\ldots,q_{L_m}))$}
	\begin{algorithmic}[1] 
		\REQUIRE Inconsistent assertions associated to $\langle\T,(q_{L_i},\ldots,q_{L_m})\rangle$     
		\ENSURE  Consistency rank
				\IF{$ \langle\T,(q_{L_i},\ldots,q_{L_m})\rangle$ is consistent}
		\RETURN $m$
		\ELSE 
		\STATE $\alpha \leftarrow i$
		\STATE $\beta\leftarrow m$
		\STATE $\gamma \leftarrow \lfloor \frac{\alpha + \beta}{2} \rfloor$
		\IF{$\langle\T,(q_{L_i} \cup \ldots \cup q_{L_{\gamma}})\rangle$ is consistent}
		\STATE $CnsRank(\T,(q_{L_\gamma},\ldots,q_{L_m}))$
		\ELSE
		\STATE $CnsRank(\T,(q_{L_i},\ldots,q_{L_\gamma}))$
		\ENDIF
		\ENDIF
	\end{algorithmic}
	\label{fc}
\end{algorithm}

\section{Assertions Associated with the Answers Profile Repairs}
\label{poslin}

This section proposes three repairs to cope with inconsistent answers that seen as a set of facts. The input of these approaches is a prioritized \textit{DL-Lite} \kb $\;$ with the prioritized profile, a query  and the profile of sets of answers. On other hand, the output of our approaches is  a consistent assertions associated to the sets of these  answers (repair answers).  
\subsection{Possibilistic-Based Repairs Answers} 
Possibility theory \cite{ref_Dubois2007} and possibilistic logic \cite{ref_Didier94} are natural frameworks to deal with uncertain, incomplete, qualitative and prioritized information. One of the interesting aspects of possibilistic KBs is the ability of reasoning with partially inconsistent knowledge \cite{ref_Dubois1991}.  As shown in \cite{ref_BenferhatBL13}, the entailment in possibilistic \textit{DL-Lite},  an adaptation of \textit{DL-Lite} entailment within a possibility theory setting, is based on the selection of one consistent, (not necessarily 
maximal) subset of $\K$. The subset $\pi(q_{P_s})$ is formed by assertions with priority levels that are less or equal to $CnsRank(q_{P_s})$.

 More formally, $\pi(q_{P_s})$ is the repair answers of $q_{P_s}$ defined by $\pi(q_{P_s}) = q_{L_1} \cup \ldots \cup q_{(CnsRank(q_{P_s}))}$. If $q_{P_s}$ is consistent with the TBox then we simply let $\pi(q_{P_s}) = q_{L_1} \cup \ldots \cup q_{L_m}$. The Algorithm \ref{algoP} returns the possibilistic-based repair answers. 
\begin{algorithm}[H]
	\caption{$\pi(\T,(q_{L_i},\ldots,q_{L_m}))$}
	\begin{algorithmic}[1] 
		\REQUIRE Inconsistent assertions associated to $\langle\T,(q_{L_i},\ldots,q_{L_m})\rangle$     
		\ENSURE Consistent assertions $\pi(q_{P_s})$

		\STATE $\alpha \leftarrow  CnsRank(\T,(q_{L_1},\ldots,q_{L_m}))$ /*Calculate recursively the consistency rank*/
		
		\RETURN $\pi(q_{P_s}) = q_{L_1} \cup \ldots \cup q_{\alpha}$   
		
	\end{algorithmic}
	\label{algoP}
\end{algorithm}

This algorithm requires $\log_2 (m)$ inconsistency tests on a set of $m$ assertions associated to $m$ answers to a query $q(x)$ with respect to $P_s$.  It returns the possibilistic based repair answers in polynomial time.

The possibilistic conclusions are considered intact since our algorithm stops in the first assertions associated to an answer $S_i$ where inconsistency is introduced. Hence, only the assertions having a degree strictly less or equal than the one the consistency degree are taken into account. However, assertions with priority levels strictly greater than the consistency degree are simply inhibited despite being unaffected by any conflict. To overcome this limitation, we improve possibilistic based repair answers  to  Linear-based repair answers.
\subsection{Linear-Based Repairs Answers}
In order to recover the assertions inhibited by possibilistic-based repair answers, we propose a new way corresponds to the use of linear-based repair answers from $q_{P_s}$. 
	\label{def6}

	Let $\K_P{=}\KB{\T}{P_s}$ be a prioritized \textit{DL-Lite} \kb $\;$ with:\\ $P_s=\{L_1,\ldots,L_m\}$, $q(x)$ be a query, $S_{P_s}=\{S_1,..., S_m\}$ a set of answers to a query $q(x)$ with respect to $P_s$, $q_{P_s} = (q_{L_1},\ldots,q_{L_m})$ is the set of assertions associated to $S_{P_s}$ such that $q_{L_i} = \{q(s) \, : \,  s \in S_{i}\}$. The linear-based repair answers of $q_{P_s}$, denoted by: $\ell(q_{P_s}) = S^{'}_1 \cup \ldots \cup S^{'}_m$ is defined as follows:
	\begin{equation} 
	S^{'}_i = 
	\left\lbrace 
	\begin{array}{r l} 
	q_{L_i} & \quad  \mbox{if} \hspace{0.1cm} \langle\T,S^{'}_1 \cup \ldots \cup S^{'}_{i-1} \cup q_{L_i}\rangle \hspace{0.5cm} \mbox{is consistent} \\ 
	\emptyset & \quad Otherwise 
	\end{array}\right. 
	\end{equation}

 $\ell(q_{P_s})$ is obtained by discarding the set of assertions $q_{L_i}$ when it conflicts with the previous set $q_{L_{i-1}}$. The following Algorithm \ref{algoL} implements the $\ell(q_{P_s})$.
\begin{algorithm}[H]
	\caption{$\ell(\T,(q_{L_1},\ldots,q_{L_m}))$}
	\begin{algorithmic}[1] 
		\REQUIRE Inconsistent assertions associated to $\langle\T,(q_{L_1},\ldots,q_{L_m})\rangle$      
		\ENSURE Consistent assertions $\ell(q_{P_s})$
		\STATE $\alpha \leftarrow CnsRank (\T,(q_{L_1},\ldots,q_{L_m}))$ /*Calculate recursively the consistency rank*/
		\STATE $\ell(q_{P_s}) \leftarrow \pi(q_{P_s})$
		\FOR{$i=\alpha +1$ to $m$ }   
		\IF{$\langle\T,\ell(q_{P_s}) \cup q_{L_i}\rangle$ is consistent}
		\STATE $\ell(q_{P_s}) \leftarrow \ell(q_{P_s}) \cup q_{L_i}$
		\ENDIF 
		\ENDFOR 
		\RETURN $\ell(q_{P_s})$
	\end{algorithmic}
	\label{algoL}
\end{algorithm}
 The time complexity of computing $\ell(q_{P_s})$ is in $P$. In fact, according to Algorithm \ref{algoL}, the computational complexity of computing $\ell(q_{P_s})$ needs $m$ executions to verify the consistency of the set of assertions $q_{P_s}$.

\subsection{\textit{Non-defeated} Repair Answers}
This new inference makes also to get a preferred repair answers. It consists in determining among the union of the set of assertions associated to the answers to a given query with respect to the ABox Profile $P_s$, the set of $\textit{free}$ elements. 
		
	Let $\K_P{=}\KB{\T}{P_s}$ be a prioritized \textit{DL-Lite} \kb $\;$ with: $P_s=\{L_1,\ldots,L_m\}$, $q(x)$ be a query, $S_{P_s}=\{S_1,..., S_m\}$ a set of answers to a query $q(x)$ with respect to $P_s$, $q_{P_s} = (q_{L_1},\ldots,q_{L_m})$ is the set of assertions associated to $S_{P_s}$ such that $q_{L_i} = \{q(s) \, : \,  s \in S_{i}\}$. We define The \textit{non-defeated} repair answers, denoted by: $nd(q_{P_s}) = S^{'}_1 \cup \ldots \cup S^{'}_m$ as follows:
	$$\forall i = 1 \ldots m, S^{'}_i = free(q_{L_1} \cup \ldots \cup q_{L_i})$$ Namely,
	$nd(q_{P_s}) = free(q_{L_1}) \cup free(q_{L_1} \cup q_{L_1}) \cup \ldots \cup free(q_{L_1} \cup \ldots \cup q_{L_m})$.

The \textit{non-defeated} repair is computed in polynomial time in \textit{DL-Lite} but  its computation is hard in propositional logic setting. In what follows, we present Algorithm \ref{algond} which computes the \textit{non-defeated} repair answers. The complexity of this algorithm is $\mathcal{O}(m)$ where $m$ is the number of answers to a query $q(x)$ \wrt the profile $P_s$.

\begin{algorithm}[H]
	\caption{$nd(\T,(q_{L_1},\ldots,q_{L_m}))$}
	\begin{algorithmic}[1] 
		\REQUIRE Inconsistent assertions associated to $\langle\T,(q_{L_1},\ldots,q_{L_m})\rangle$       
		\ENSURE Consistent assertions $nd(q_{P_s})$
		
		\STATE $\alpha \leftarrow CnsRank (\T,(q_{L_1},\ldots,q_{L_m}))$ /*Calculate recursively the consistency rank*/
		
		\STATE $nd(q_{P_s}) \leftarrow \pi(q_{P_s})$
		\FOR{$i=\alpha+1$ to $m$} 
		
		\STATE $nd(q_{P_s}) \leftarrow nd(q_{P_s}) \cup free(q_{L_1} \cup \ldots \cup q_{L_i})$
		
		\ENDFOR 
		
		\RETURN $nd(q_{P_s})$
	\end{algorithmic}
	\label{algond}
\end{algorithm}

In the following example, we show that the productivity of our repairs answers strategies applied on prioritized KB after runing querying is more than the repairs strategies applied on the whole KB before querying. We mean by productivity, the number of answers returned by applying each algorithm.

	\label{EXm}
	Let  $\K= \langle \T, P_{s} \rangle$  be a prioritized \textit{DL-Lite} knowledge base. Then we have:
	
	\noindent $\T = \{A \isa \; \neg B, A \isa \;\neg E, E \isa D , R \isa P \}$ and $P_s =\{ L_1, L_2, L_3, L_4, L_5\}$ such that:
	
	\begin{center}
		\begin{tabular}{|l|}
			\hline
			$L_1$ = $\{ A(a), R(a,z), A(c)\}$,\\
			$L_2$ = $\{ B(a), R(b,z),A(b)\}$,\\
			$L_3$ = $\{ B(a),R(a,z), B(c)\}$,\\
			$L_4$ = $\{ E(e), R(e,z), A(c)\}$,\\
				$L_5$ = $\{A(e), R(e,z), A(c), R(c,z) \}$\\
			\hline
		\end{tabular}
	\end{center}
Let we have the following query (giving all concepts $x$ which are on relation with $z$):\\   $q(x)= \exists x. R(x,z)$. 
	We get the following set of assertions $q_{P_s}$:\\
	\begin{center}
		\begin{tabular}{|l|}
			\hline
			$q_{L_1} = \{A(a)\}$,\\
			$q_{L_2} = \{A(b)\}$,\\
			$q_{L_3} = \{B(a)\}$,\\
			$q_{L_4} = \{E(e)\}$,\\
			$q_{L_5} = \{A(e), A(c)\}$\\
			\hline
		\end{tabular}
	\end{center}
	One can check that the set of conflictsis  answers: 
	\begin{center}
		\begin{tabular}{|l|}
			\hline
			$ \mathcal{C}(qp_{s}) =\{(A(a), B(a)); (A(e),E(e))\}$\\
			\hline
		\end{tabular}
	\end{center}
	
	However,  the set of conflicts is of KB (according to \cite{ref_BenferhatBT15,ref_TelliBBBT17}):
	\begin{center}
		\begin{tabular}{|l|}
			\hline
			$\mathcal{C}(P_{s})=\{(A(a), B(a)); (A(e), E(e)); (A(c), B(c))  \}$\\
			\hline
		\end{tabular}
	\end{center}
	
	In addition, the set of $free$ elements for the assertion $qp_{s}$ is:
	
	\begin{center}
		\begin{tabular}{|l|}
			\hline
			$ free(qp_{s})= \{A(b), A(c) \}$\\
			\hline
		\end{tabular}
	\end{center}
	
	While that, the set $free$ elements for $P_{s}$ is:
	
	\begin{center}
		\begin{tabular}{|l|}
			\hline
			$ free(P_{s})= \{R(a,z), R(b,z), A(b), R(e,z), R(c,z) \}$\\
			\hline
		\end{tabular}
	\end{center}
	
	According to Algorithm \ref{fc}, the consistency rank of $qp_{s}$ equal 2. While,  the consistency rank of $P_{s}$ equal 1. Moreover, we have the the following free sets of $qp_{s}$:

	\begin{center}
		\begin{tabular}{|l|}
			\hline
			$free(q_{L_{1}}) = \{A(a)\}$,\\
			$free(q_{L_{1}} \cup q_{L_{2}})= \{A(a), A(b)\}$,\\
			$free(q_{L_{1}} \cup ... \cup q_{L_{3}} ) = \{A(a), A(b)\}, $\\
			$free(q_{L_{1}} \cup ... \cup q_{L_{4}} ) = \{A(a), A(b), E(e)\} $,\\
			$free(q_{L_{1}} \cup ... \cup q_{L_{5}} ) = \{A(a), A(b), E(e), A(c)\} $\\
			\hline
		\end{tabular}
	\end{center}

Using the definitions of possibilistic-based repair answers, linear-based repair answers and \textit{non-defeated} repair answers,  we have:
	\begin{center}
		\begin{tabular}{|l|}
			\hline
			$\pi(q_{P_s}) = \{A(a), A(b)\}$,\\
			$\ell(q_{P_s})= \{A(a), A(b), E(e)\}$,\\
			$nd(q_{P_s}) = \{A(a), A(b), E(e), A(c)\} $\\
					\hline
		\end{tabular}
	\end{center}
	
		Now, we have the the following free sets of $P_{s}$:

	\begin{center}
	\begin{tabular}{|l|}
		\hline
		$free(L_{1}) = \{A(a), R(a,z), A(c)\}$,\\
		$free(L_{1} \cup L_{2})= \{A(a), R(a,z), A(c), R(b,z), A(b)\}$,\\
		$free(L_{1} \cup ... \cup L_{3}) = \{A(a), R(a,z), A(c), R(b,z), A(b)\}, $\\
		$free(L_{1} \cup ... \cup L_{4}) = \{A(a), R(a,z), A(c), R(b,z), A(b), E(e), R(e,z) \} $,\\
		$free(L_{1} \cup ... \cup L_{5}) = \{A(a), R(a,z), A(c), R(b,z), A(b), E(e), R(e,z), R(c,z)\} $\\
		\hline
	\end{tabular}
\end{center}

	Although, if we use the definitions of possibilistic-based repair, linear-based repair and  \textit{non-defeated} repair  proposed in (\cite{ref_BenferhatBT15,ref_TelliBBBT17}) directly on  $P_{s}$ (before querying), we have:
	
	\begin{center}
		\begin{tabular}{|l|}
			\hline
			$\pi(P_{s})= \{A(a), R(a,z), A(c)\}$,\\
			$\ell(P_{s})= \{A(a), R(a,z), A(c), R(e,z), E(e)\}$,\\
			$nd(P_{s})= \{A(a), R(a,z), A(c), R(e,z), E(e), R(c,z)\}$\\
		\hline
		\end{tabular}
	\end{center}
	
	By applying the same query $q(x)$ on these repairs, we have:
	
	\begin{center}
		\begin{tabular}{|l|}
			\hline
			$\pi(P_{s)} \, \models \, q$ = $\{A(a)\}$,\\
			$\ell(P_{s}) \, \models \, q$ = $\{A(a), E(e)\}$,\\
			$nd(P_{s}) \, \models \, q$ = $\{A(a), A(c), E(e)\} $\\
						\hline
		\end{tabular}
	\end{center}
	
	Clearly, $\pi(P_s) \; \models \; q \subseteq \pi(q_{P_s})$, $\ell(P_s) \; \models \; q \subseteq \ell(q_{P_s})$, $nd(P_s) \; \models \; q \subseteq nd(q_{P_s})$. \finexample

Next section is an experimental study which based on recursive programming (Consistency Rank function) about the running time of our approach.
\section{Experimental Evaluation}
\label{exp}
We implemented our proposed algorithms in Java for computing a repair answering in prioritized assertional bases under inconsistent \kb s. The parses \textit{\dllite{}} \kb s expressed in \emph{OWL2-QL} function syntax and a SQLite database engine. We used a part of benchmark \footnote{Available at: \url{https://code.google.com/p/combo-obda/}} we considered the LUBM$^{\exists}{20}$ ontology (\ie{} TBox) \cite{ref_lutz13}, and we generated by the Extended University Data Generator (EUDG) an ABox   contains $1000$ assertions and  we split them into 5 strata. These ABoxes  with respectively  \emph{50}, \emph{200} and \emph{500} conflict sets. We ran the proposed algorithms in \cite{ref_TelliBBBT17} and our developed algorithms for computing repairs before and after launching a instance, ground and conjunctive query.




We are interested in the basic performance metrics used for evaluate our algorithms. In our case, our system classifies the assertions of the ABox into two classes: consistent and inconsistent. Consistent assertions are placed by the system in the positive class, and inconsistent assertions are placed by the system in the negative class.

 When our algorithm classification is correct, the assertions are retrieved. However, if the algorithm makes a mistake, the assertions are not retrieved. We can compute the four following performance indices:

\begin{itemize}
	\item CR is the number of consistent retrieved assertions after applying the repairs.
	\item CNR is the number of consistent  not retrieved assertions after applying the repairs.
	\item IR is the number of inconsistent retrieved assertions before applying the repairs.
	\item INR is the number of inconsistent  not retrieved assertions before applying the repairs.
\end{itemize}

In the following, we present the precision, recall and F-measure measures (\cite{ref_daglib,ref_RaghavanJB89}), which we will use to evaluate the performance of our algorithms:

\begin{itemize}
	\item Precision (P) is the ratio of the number of consistent assertions retrieved by the total number of retrieved assertions.
	
The principle of precision measure: when we ask a query on our ABox, we wish that the assertions proposed as answer correspond to our expectations. All retrieved irrelevant assertions constitute what is called "the noise". Precision, is opposed by assertional noise. If it is high, this indicates that few unnecessary assertions are offered by the system and that the system can be considered "precise".
	\begin{equation}
	P = \frac{CR}{CR + IR}
	\end{equation}
	\item Recall (R) is the ratio of the number of consistent assertions retrieved by the total number of consistent assertions.
	
The principle of recall measure: when we ask a query on our ABox, we wish to see appearing all assertions that could answer our need of information. If this correspondence between the questioning of the user and the number of assertions presented is important then the recall rate is high. Conversely, if the system have many interesting assertions but they do not appear in the list of answers, we speak of "silence". Silence opposes recall.
	\begin{equation}
	R = \frac{CR}{CR + CNR}
	\end{equation}
	\item F-measure (F) is the harmonic average of the precision P and the recall R:
	\begin{equation}
	F = \frac{2*(P * R)}{(P + R)}
	\end{equation}
\end{itemize}

The following table shows the precision, recall and F-measure measures for our algorithms, after launching a instance, ground and conjunctive query on our ABox.
\begin{center}
	\begin{small}
		\begin{tabular}{|c|c|c|c|c|c|c|c|c|c|c|} 
			\hline
			Conflict & Query & \multicolumn{3}{c}{$\pi(q_{P_s})$} & \multicolumn{3}{|c|}{$\ell(q_{P_s})$} & \multicolumn{3}{c|}{$nd(q_{P_s})$} \tabularnewline
			\cline{3-11} size & type & $P $ & $R $ & $F $ & $P $ & $R $ & $F $ & $P $ & $R $ & $F $ \tabularnewline 
			\cline{3-11}
			\hline
			\multirow{3}{*}{\textbf{50}} & Instance & 84.03 & 94.33& 88.88 & 84.67& 94.59& 89.35& \textbf{88.41}& \textbf{96.02}& \textbf{92.05} \tabularnewline
			
			& Ground & 79.78 & 92.59 & 85.70& 80.80 & 93.02 & 86.48& \textbf{87.24} & \textbf{95.58} & \textbf{91.21 } \tabularnewline
			
			& Conjunctive & 75.94 & 90.90 & 82.74 & 77.38 & 91.54 & 83.86& \textbf{78.65} & \textbf{92.10} & \textbf{84.84}  \tabularnewline
			\hline
			\multirow{3}{*}{\textbf{200}} & Instance  & 55.55 & 83.33 & 66.66 & 55.55 & 83.33& 66.66& \textbf{60.97} & \textbf{86.20}& \textbf{71.42} \tabularnewline
			
			& Ground & 52.94& 81.81 & 64.28& 52.94& 81.81 & 64.28& \textbf{55.55} & \textbf{83.33}& \textbf{66.66 } \tabularnewline
			
			&  Conjunctive & 42.85 & 75 & 54.53& 42.85 & 75 & 54.53& \textbf{51.51} & \textbf{80.95}& \textbf{62.95  } \tabularnewline
			\hline    
			\multirow{3}{*}{\textbf{500}} & Instance & 24.24 & 48.97 & 32.42& 24.24 & 48.97 & 32.42& \textbf{36.97} & \textbf{63.76}& \textbf{46.80}  \tabularnewline
			
			&  Ground & 21.05 & 44.44 & 28.56& 21.05 & 44.44 & 28.56& \textbf{34.78} & \textbf{61.53}& \textbf{44.44} \tabularnewline
			
			& Conjunctive & 21.05 & 44.44 & 28.56& 21.05 & 44.44 & 28.56& \textbf{32.43} & \textbf{59.01}& \textbf{41.85} \tabularnewline
			
			\hline
		\end{tabular}
	\end{small}
	
	\captionof{table}{Experimental evaluation of proposed inferences expressed in \%}
\end{center}
 All retrieved inconsistent assertions constitute what is called "the noise". Precision, is opposed by assertional noise. If it is high, this indicates that few unnecessary assertions are offered by the system and that the system can be considered "precise".

The precision measure obtained in the previous table show that the non-defeated algorithm is more precise than the two others. Also, when increasing the number of conflicts, the precision measure decreases in the three algorithms. Hence, this measure is influenced by the number of conflicts in the ABox.

Now, if the correspondence between the questioning of the user and the number of assertions presented is important, then the recall rate is high. Conversely, if the system have many interesting assertions but they do not appear in the list of answers, we speak of "silence" (silence opposes recall).

Similarly, according to the results of the previous table, we note that the recall measure of the non-defeated algorithm is higher than the other algorithms. In addition, when increasing the number of conflicts, the recall measure decreases with the three algorithms. Thus, this measure is influenced by the number of conflicts in the ABox.

Now, we are interested in the time taken to compute our proposed algorithms of repairs. For this aim, we generated and   splitted the ABoxs respectively into 3 strata, 5 strata and then 7 strata. These ABoxes contain respectively 50, 200 and 500 conflict sets. The results of this exprentation are shown in Table \ref{t2}.

\begin{small}
\begin{center}
	\begin{tabular}{|c|c|c|c|c|c|c|c|} 
		\hline
		
		Conflict & Strata &\multicolumn{3}{c}{Before querying (\cite{ref_BenferhatBT15,ref_TelliBBBT17})}&\multicolumn{3}{|c|}{After querying}  
		\tabularnewline
		\cline{3-8} size & level & $\pi({P_s})$ & $\ell({P_s})$& $nd({P_s})$&  $\pi(q_{P_s})$ & $\ell(q_{P_s})$& $nd(q_{P_s})$  \tabularnewline 
		\cline{3-8}
		\hline
		\multirow{3}{*}{50} & 3 & 10.93 & 11.09& 45.45  & 2.83& 2.88& 11.50  \tabularnewline
		
		& 5 & 10.96 & 32.65 & 67.43 & 2.85 & 8.22 & 17.50  \tabularnewline
		
		& 7 & 16.36 & 32.92 & 94.26 & 4.10 &  8.25 & 23.58 	\tabularnewline
		\hline
		\multirow{3}{*}{200} & 3  & 12.20
		& 28.17
		& 47.90 & 3.06 & 3.80& 11.96 \tabularnewline
		
		& 5 & 12.47& 33.09
		& 78.18	&3.12 & 8.25& 19.55  \tabularnewline
		
		&  7 & 16.53
		& 38.53 & 96.99 & 4.17 & 9.65& 24.25 \tabularnewline
		\hline    
		\multirow{3}{*}{500} & 3 & 13.86
		& 14.41 & 50.68& 3.45 &7.07 & 12.77 \tabularnewline
		
		&  5 & 18.12 & 29.98 & 93.36& 4.55 & 8.27& 23.50 \tabularnewline
		
		& 7 & 19.89
		& 43.41 & 99.43 &4.98 & 10.90& 25.00 \tabularnewline
		
		\hline
		
	\end{tabular}
	\captionof{table}{Runing time of our repairs before and after querying  (in seconds)}
	\label{t2}
\end{center}
\end{small}

As expected,  the existing strategies which based on access to the whole Knowledge Base take more running time than our proposed strategies which which handle only with the set of answer of a specific request. Precisely, the computing of our repairs after querying requires in most of our experiments, less time than its computing before querying. However, the time needed for computing the non-defeated repair answers increases with the size of conflicts in the ABox.
Finally, according to the results obtained, we conclude that non-defeated algorithm is the most performing followed by linear and possibilistic algorithms.

Now, we are interested to evaluate the productivity of our repair algorithms before and after querying. We  mean by productivity, the assertions that are preserved from the ABox (\resp \; answers) in order to restore the consistency of the \kb $\,$ (\resp \; answers)

\begin{small}
\begin{center}
\begin{tabular}{|c|c|c|c|c|c|c|c|} 
\hline

Conflict & Query &\multicolumn{3}{c}{Before querying}&\multicolumn{3}{|c|}{After querying}  
 \tabularnewline
\cline{3-8} size & type & $\pi(P_s) \; \models \; q$ & $\ell(P_s) \; \models \; q$& $nd(P_s) \; \models \; q$& $\pi(q_{P_s})$ & $\ell(q_{P_s})$& $nd(q_{P_s})$ \tabularnewline 
\cline{3-8}
\hline
\multirow{3}{*}{50} & Instance & 15 & 16& 25 & 20& 21& 29 \tabularnewline
 
 & Ground & 14 & 15 & 24& 15 & 16 & 26  \tabularnewline
                   
  & Conjunctive & 10 & 12 & 14 & 12 & 13 & 14  \tabularnewline
 \hline
\multirow{3}{*}{200} & Instance  & 18 & 19 & 22 & 20 & 20& 25 \tabularnewline

           & Ground & 17& 18 & 20& 18 & 18& 20  \tabularnewline
                     
          &  Conjunctive & 10 & 10 & 15& 12 & 12& 17   \tabularnewline
\hline    
\multirow{3}{*}{500} & Instance & 11 & 11 & 20& 12 & 12& 22  \tabularnewline

         &  Ground & 11 & 11 & 18& 10 & 10& 20 \tabularnewline
                     
        & Conjunctive & 10 & 10 & 17& 10 & 10& 18
          \tabularnewline
                    
\hline

\end{tabular}
\captionof{table}{Productivity of applying our repair algorithms before and after querying (expressed in \%)}
\end{center}
\end{small}

 \begin{figure}[H] 
\centering
\includegraphics[width=9cm]{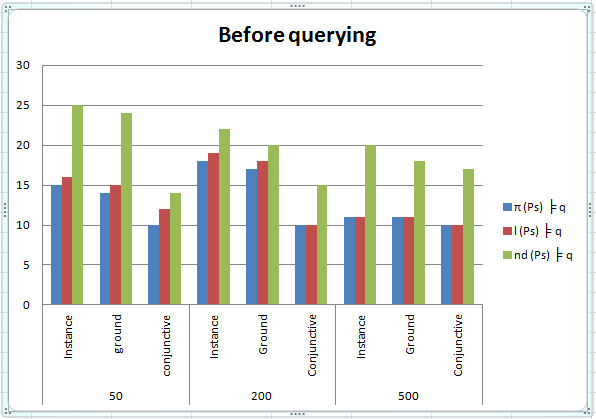} 
\caption{Productivity of applying our repair algorithms before querying} 
\end{figure}


 \begin{figure}[H] 
\centering
\includegraphics[width=9cm]{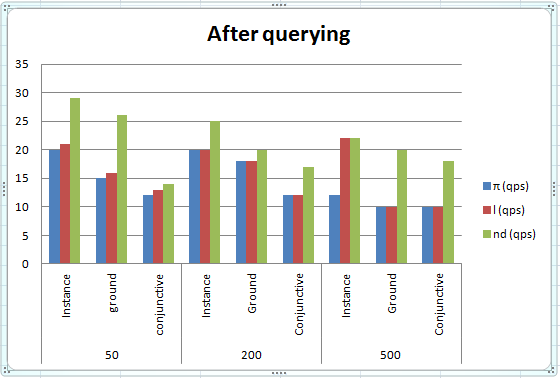} 
\caption{Productivity of applying our repair algorithms after querying} 
\end{figure}


From Table 3, figure 1 and figure 2, the productivity of possibilistic-based repair answers is very cautious comparing to the other strategies.  Namely, for a given ABox and a given number of strata, possibilistic-based repair answers has the  largest number of dropped elements. 
 This similarly holds for the linear-based repair answers when there exists at least a conflict in each strata.  Hence, it is obvious that the size of  conflicting elements in the assertional bases is one of  main parameters that influence the productivity of the repairs answers. The \textit{non-defeated} repair answers gives a significant number of consistent answers compared to the other strategies.
 
Table 1 shows that the productivity of $\pi(q_{P_s})$, $\ell(q_{P_s})$ and $nd(q_{P_s})$ are more productive than  $\pi(P_s) \; \models \; q$, $c\ell(P_s) \; \models \; q$ and $nd(P_s) \; \models \; q$ respectively. We note that the computing on our proposed algorithms requires a polynomial running time.

\section{Related Works}
\label{rw}

The principal inspiration for the present paper comes from
a line of research in inconsistency-handling. Inconsistency is defined with respect to some assertions that contradict the terminology. Typically, a TBox is usually verified and validated while the assertions can be provided in large quantities by various and unreliable sources and may contradict the TBox. Moreover, it is often too expensive to manually check and validate all the assertions. This is why it is very important in OBDA (Ontology-based Data Access) to reason in the presence of inconsistency. Many works (\eg \, \cite{ref_LemboLRRS10,ref_bienvenuAAAI12}), basically inspired by database approaches (\eg \,  \cite{ref_Bertossi2011}), tried to deal with inconsistency in \textit{DL-Lite} by adapting several inconsistency-tolerant inference methods. In many applications, assertions are often provided by several and potentially conflicting sources having different reliability levels.

Moreover, a given source may provide different sets of uncertain assertions with different confidence levels. Gathering such sets of assertions gives a prioritized or a stratified assertional base. The role of priorities in handling inconsistency is very important and it is largely studied in the literature within propositional logic setting (\eg \, \cite{ref_Baral457,ref_Benferhat95}). Several
works (\eg \, \cite{ref_MartinezPPSS08,ref_StaworkoCM12,ref_du13}) studied the notion of priority when querying inconsistent databases or DL KBs. Unfortunately, in the
OBDA setting, there are only few works, such as the one given in \cite{Bienvenu2014} for dealing with reasoning under prioritized \textit{DL-Lite} ABox.

A recent line of work studies the inconsistency in lightweight ontologies. For instance,  the authors in \cite{ref_lemboLRRS15,ref_LemboLRRS10,ref_bienvenuAAAI12,ref_Benferhat2016NonObjectionIF,ref_BagetBBCMPRT16a}  investigate the problem of inconsistency in \kb s by computing a set of consistent subsets of assertions called repairs, which recovers the consistency with respect to the ontology, and then using them to answer the queries. Moreover, the authors propose in \cite{ref_BenferhatBT15,ref_TelliBBBT17} polynomials algorithms for select a single preferred repair from a prioritized inconsistent \textit{DL-Lite} \kb $\;$ to allow an efficient query answering  once the repair is computed. Particularly, the authors in \cite{ref_BenferhatBT15} propose a new approach based on the selection of only one preferred repair.  However, in \cite{ref_TelliBBBT17}, the authors propose a sequential inference strategies based on the selection of one consistent assertional base. The authors in \cite{ref_BoughammouraO17,ref_BoughammouraOH12,ref_BoughammouraHO15} propound a new algorithm which makes easy to query answering without access to Web databases. 

In our work,  a recursive algorithm starts by querying all the whole  knowledge base which will allow us to have an exhaustive list of all the possible answers. Then,  if this list of answers is inconsistent, the algorithm   repairs it. Hence, it do not any correction except if the set answers is inconsistent. However, existing algorithms start by repartion the   knowledge base, then, querying this reparation with a sequential function.
\section{Conclusion}
\label{conclusion}
We focused in this work on the problem of inconsistency answers  over  prioritized \textit{DL-Lite} Knowledge Bese. For this purpose, we started out by giving some bases notions about:  prioritized \textit{DL-Lite} knowledge bese,  inconsistency tolerant reasoning with the answers set. Then, we developed a recursive function to calculate the consistency rank in order to use it on our  proposed algorithms. The main contribution of this paper is how to repair the set of answers instead of the whole Knowledge Base without increasing the computational complexity time? The experimental studies evaluated the productivity and rapidity of running of our proposed repairs answers using the basic performance metrics: Precision, Recall and F-measure and the running time.\\
A future work is to apply our approaches to query the closure of KB in presence of possible answers.

\end{document}